\documentclass[a4paper,twocolumn,11pt, unpublished]{quantumarticle}
\pdfoutput=1

\usepackage[utf8]{inputenc}
\usepackage[english]{babel}
\usepackage[T1]{fontenc}
\usepackage{amsmath,amssymb,amsthm,thmtools}
\usepackage{mathtools,mathrsfs}
\usepackage{amsbsy}
\usepackage{hyperref}
\usepackage{tikz}
\usepackage{braket}
\usepackage{tikz-cd}
\usepackage{qcircuit}
\usepackage{float}
\usepackage[caption = false]{subfig}

\graphicspath{{./Figures/}}

\newtheorem{theorem}{Theorem}

\begin{document}

\title{A Bottom-up Approach to Constructing Symmetric Variational Quantum Circuits}

\author{B.M. Ayeni}
\email{babatunde.ayeni@mu.ie}
\affiliation{Department of Theoretical Physics, National University of Ireland, Maynooth, Ireland}

\orcid{0000-0003-4035-149X}  


\maketitle

\begin{abstract}
In the age of noisy quantum processors, the exploitation of quantum symmetries can be quite beneficial in the efficient preparation of trial states, an important part of the  variational quantum eigensolver algorithm. The benefits include building quantum circuits which are more compact, with lesser number of paramaters, and more robust to noise, than their non-symmetric counterparts. Leveraging on ideas from representation theory we show how to construct symmetric quantum circuits. Similar ideas have been previously used in the field of tensor networks to construct symmetric tensor networks. We focus on the specific case of particle number conservation, that is systems with U(1) symmetry. Based on the representation theory of U(1), we show how to derive the particle-conserving exchange gates, which are commonly used in constructing hardware-efficient quantum circuits for fermionic systems, like in quantum chemistry, material science, and condensed-matter physics. We tested the effectiveness of our circuits with the Heisenberg XXZ model.
\end{abstract}

\section{Introduction}
Despite the impressive advancements seen in quantum computing technology in just a few decades, current quantum processors are still limited in scale and with noise, hence also limiting their applications. Current and near-term quantum processors are also known as ``noisy intermediate-scale quantum" (NISQ) computers.\cite{preskill2018quantum} 

One important problem that is suitable for NISQ processors is calculating the ground state of many-particle systems in either quantum chemistry, material science, or quantum many-body physics. This involves building a quantum circuit to prepare a trial state, which is then optimized to the ground state of the problem with the help of some quantum algorithm. An approach that is currently feasible for NISQ devices is the variational quantum eigensolver (VQE) algorithm.\cite{peruzzo2014variational} This is a hybrid quantum-classical algorithm, where the trial state is  prepared as a parameterized quantum circuit on a quantum processor, and the optimization is performed on a classical computer.\cite{cerezo2021variational,mcclean2016theory,bharti2022noisy} 

However, it has been discovered that in the limit of a high number of qubits and hence exponentially large Hilbert space, VQE can suffer from ``barren plateaus.''\cite{mcclean2018barren,wang2021noise} {These are regions in high-dimensional parameter space where the gradient becomes extremely flat or close to zero during a training process. This leads to slow or stalled learning, and thus rendering variational approaches ineffective. Several approaches have been developed to suppress the effect of this problem, includcing using a local rather than global cost function,\cite{cerezo2021cost} compiling deep quantum circuits into shallow ones,\cite{robertson2022escaping,robertson2023approximate}, and several other techniques.\cite{grant2019initialization,sack2022avoiding,patti2021entanglement}}

It is well known in physics that the symmetries in a system can be used to reduce the number of parameters needed to describe the system. This remains applicable to quantum computation too, where the exploitation of symmetries can be used to reduce the number of parameters in variational quantum circuits (VQCs). Moreover, it makes symmetries advantageous in the context of NISQ computing. The idea of exploiting symmetries in quantum circuits is reviewed in Ref.~\cite{lacroix2023symmetry}. For the specific case of conserving global particle number in quantum circuits, which is also the focus of our paper, a number of works have been carried out \cite{barkoutsos2018quantum,arrazola2022universal,gard2020efficient} The basic idea underlying all these works is that the basic local gate on $k$ qubits is constructed as a (sub)circuit that conserves particle number.

In this work, we also show how to construct symmetric variational quantum circuits but through a route different from previous considerations. For inspiration, we looked to symmetric tensor networks, due to the correspondence between tensor networks and quantum circuits. The subject area of tensor networks is one where the exploitation of symmetries has been well developed, and which resulted in a plethora of works by many authors in around the past two decades.\cite{singh2010tensor,singh2011tensor,singh2012tensor,bauer2011implementing,mcculloch2002non,weichselbaum2012non,pfeifer2010simulation,konig2010anyonic,pfeifer2015finite,singh2014matrix,ayeni2016simulation,schmoll2020programming} Symmetries can be exploited in quantum circuits for the same purpose for which they are exploited in tensor networks, namely to reduce the number of parameters and/or to necessarily enforce the preservation of certain symmetries when studying systems with symmetry-protected phases. In addition, the use of symmetries can be important for NISQ devices, as they are noisy, and hence can be used to reduce error in calculations by ``locking'' the quantum circuit in a relevant symmetry sector of the Hilbert space.

In our approach, we show how to construct symmetric VQCs that conserve total particle number (or equivalently U(1) symmetry). Our constructions are based on ideas from the representation theory of Abelian groups, such as the continuous group U(1) and the finite cyclic group $\mathbb{Z}_k$. We showed how to construct elementary symmetric variational quantum gates that can then be composed to build desired symmetric variational quantum circuits. The methods we expose can also be generalized to handle other types of quantum symmetry.

The rest of the paper is structured as follows: In Sec.~\ref{Sec: symmetric states and operators}, we set the stage by recalling the definitions of symmetric many-body quantum states and symmetric operators. We especially considered how a U(1) symmetry defined in terms of particle number is encoded into a particle-conserving matrix operator, and in general any tensor. In Sec.~\ref{Sec:Symmetric circuit elements}, we present how to systematically build symmetric variational quantum gates. In Sec.~\ref{Sec:Number Conserving Circuit}, we show how to use the symmetric quantum gates to construct particle-conserving quantum circuits. In Sec.~\ref{Sec:Test Model}, we test the circuits on the 1D Heisenberg XXZ model, a model that has U(1) symmetry. In Sec.~\ref{Sec:Conclusion}, we conclude the paper and present an outlook for further research.

\section{Symmetric many-body states and operators}
\label{Sec: symmetric states and operators} 
In this section, we start by recalling basic definitions of symmetric many-body quantum states and symmetric many-body operators. First, in the general context, and secondly we specialize to U(1) symmetry, which can be associated with particle number conservation.

\subsection{General consideration}
Consider a one-dimensional lattice $\mathcal{L}$ with $L$ number of sites, with the vector space $\mathbb{V}^{\otimes L}$. An arbitrary pure state in this space can be written in a tensor-product basis as
\begin{equation}
\label{Eq:PureState}
    \ket{\psi} = \sum_{i_1, i_2, \ldots, i_L} c_{i_1, i_2, \ldots, i_L} \ket{i_1, i_2, \ldots, i_L},
\end{equation}
where the complex coefficient $c_{i_1, i_2, \ldots, i_L}$ is the probability amplitude for each product basis state. The collection of all the coefficients can be regarded as an $L$-index tensor. The number of coefficients scales exponentially as $2^L$.

\emph{Invariant states.} 
A many-body quantum state $\ket{\Psi} \in \mathbb{V}^{\otimes L}$ with a global, onsite symmetry is invariant with respect to the action of a group $\mathcal{G}$ if 
\begin{equation}
    \left(\hat{U}_g\right)^{\otimes n} \ket{\Psi} = \ket{\Psi}, \qquad \forall g \in \mathcal{G},
\end{equation}
where $\hat{U}_g: \mathbb{V} \rightarrow \mathbb{V}$ is the unitary representation of $g \in \mathcal{G}$ on each site.

The implication of the symmetric constraint on $\ket{\Psi}$ (Eq.~\eqref{Eq:PureState}) is that the coefficients transform as
\begin{equation}
\label{Eq:Symmetric quantum state equation}
    c_{i'_1, \ldots, i'_L} =  \sum_{i_1, i_2, \ldots, i_L} (\hat{U}_g)_{i'_1, i_1} \ldots (\hat{U}_g)_{i'_L, i_L}   c_{i_1, \ldots, i_L},
\end{equation}
where it can be seen how the operators, $\hat{U}_g$, act on the indices of the tensor $c_{i_1, \ldots, i_L}$.

\begin{figure}
    \centering
    \includegraphics[width=\columnwidth]{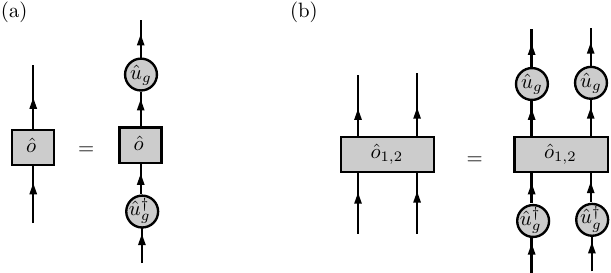}
    \caption{Examples of one-site and two-site symmetric operators with respect to a group action on them: (a) On a single site $\hat{o} = \hat{U}_g \hat{o} \hat{U}_g^{\dagger}$. (b) On two sites $\hat{o}_{1,2} = \left(\hat{U}_g \otimes \hat{U}_g \right) \hat{o}_{1,2} \left(\hat{U}_g^{\dagger} \otimes \hat{U}_g^{\dagger} \right) $. The ``orientation'' of the indices of the tensors are delineated with arrows on the legs of their diagrams, where $\hat{U}_g$ can be said to act on ``outgoing indices'' and $\hat{U}_g^{\dagger}$ acts on ``incoming indices.''} 
    \label{Fig:symm_operators}
\end{figure}

\emph{Symmetric operators.} An operator $\hat{O} : \mathbb{V}^{\otimes n} \rightarrow \mathbb{V}^{\otimes n}$ is symmetric with respect to the action of the group $\mathcal{G}$ if 
\begin{equation}
    \left[ \hat{U}_g^{\otimes n}, \hat{O} \right] = 0,
\end{equation}
or equivalently that 
\begin{equation}
    \hat{O} = \left(\hat{U}_g\right)^{\otimes n} \hat{O} \left(\hat{U}_g^{\dagger} \right)^{\otimes n}.
\end{equation}
In terms of components, this is 
\begin{align}
O^{i''_1 \ldots i''_n}_{i'''_1 \ldots i'''_n} = (\hat{U}_g)_{i''_1, i'_1} & \ldots  (\hat{U}_g)_{i''_n, i'_n}  O^{i'_1 \ldots i'_n} _{i_1 \ldots i_n} \nonumber \\
& \times (\hat{U}_g)^{\dagger}_{i_1, i'''_1} \ldots (\hat{U}_g)^{\dagger}_{i_n, i'''_n}.
\end{align}
Examples for one and two sites are shown in Fig.~\ref{Fig:symm_operators}\footnote{We shall employ the common tensor diagrammatic notations to sometimes denote tensors, for convenience.}. Though this applies in general to any $n$-body operators, for example Hamiltonians (where $n=L$), (reduced) density matrices, etc. Note that the operators  can be considered as tensors having the same number of ``incoming'' and ``outgoing'' indices. In addition, each local vector space of the tensor product has the same dimension and hence transforms by the same representation.

\emph{Symmetric tensors.} An arbitrary symmetric tensor can have any number of incoming and outgoing indices with possibly different dimensions, and therefore the symmetry group acts on different indices (or legs) with possibly different unitary representations. We provide an example that illustrates the idea. Assume a linear operator, $T$ acts as
\begin{equation}
    T : \mathbb{C}^m \rightarrow \mathbb{C}^{m_1} \otimes \mathbb{C}^{m_2}, 
\end{equation}
where the dimensions $m, m_1$, and $m_2$ are assumed to be unequal. The linear operator can be written as 
\begin{equation}
    T = \sum_{\alpha, \beta, \gamma} T_{\gamma}^{\alpha  \beta} \ket{\alpha \beta} \bra{\gamma},
\end{equation}
where $1 \leq \gamma \leq m$, etc. Let the representations of $g \in \mathcal{G}$ acting on the different indices of $T$ be $\hat{u}_g$, $\hat{v}_g$, and $\hat{w}_g$. The tensor $(T_{\gamma}^{\alpha  \beta})$ transforms as
\begin{equation}
T^{\alpha',\beta'}_{\gamma'} = \sum_{\alpha,\beta, \gamma} (\hat{v}_g)_{\alpha' \alpha} (\hat{w}_g)_{\beta' \beta}  T^{\alpha,\beta}_{\gamma} (\hat{u}_g^{\dagger})_{\gamma \gamma'}.     
\end{equation}
This can be represented with a diagram as 
\begin{equation}
\begin{matrix}
      \includegraphics[width=0.5\columnwidth]{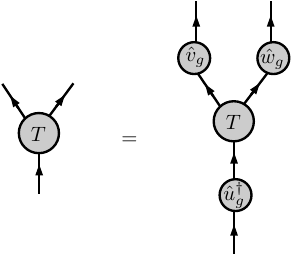} 
\end{matrix},
\end{equation}
where $\hat{u}^{\dagger}_g$ acts on the ``incoming'' leg, and $\hat{v}_g$ and $\hat{w}_g$ act on the ``outgoing'' legs. 

This can be generalized to a tensor of any order, and with any number of incoming and outgoing indices.

\subsection{Specialization to  U(1) symmetry}
\label{Sec:U1Tensors}
In the light of the general discussion of symmetry in many-body quantum systems in the previous section, which is true for any symmetry group, we now focus specifically on the group U(1).\footnote{The group $U(1)$ is the set of all $1 \times 1$ complex matrices, which is isomorphic to the set of unit complex numbers $\{ e^{i\theta}, ~ \theta \in [0, 2\pi) \}$.} We shall examine the constraints it imposes on the structure of symmetry-invariant  tensors.\cite{singh2011tensor} The generator of the group U(1) can be associated with a particle number operator. 

\emph{U(1)-symmetric quantum state}. A many-body state on a lattice of $L$ sites with a U(1) symmetry has a global conserved particle number, say $N$. The quantum state can be written as
\begin{equation}
    \ket{\Psi_N} = \sum_{\mu_N} \psi_{N \mu_N} \ket{N \mu_N},
\end{equation}
where $\{\ket{N \mu_N}  \}$ is the basis in the relevant N-particle subspace $\mathbb{V}_N$,\footnote{The Hilbert space $\mathbb{V}^L$ of a lattice of $L$ sites can be written in terms of Fock space as
\begin{equation}
    \mathbb{V}^{\otimes L} = \bigoplus_{n=0}^L \mathbb{V}_n.
\end{equation}
Therefore, an N-particle symmetric quantum state is restricted to live in the N-particle sector, $\mathbb{V}_N$, of the Hilbert space.} and $N=0, 1, \ldots, L$ (assuming occupancy of a maximum of one particle per site). The basis $\{\ket{N \mu_N}  \}$ is called the \emph{charge-degeneracy} (CD) basis, where $N$ is the global conserved charge and $\mu_N$ is the degeneracy index of the charge. Note that, like the degeneracy of eigenstates of a physical system, the \emph{degeneracy} used here refers to the total number of states having the same total number of particles. This basis can also be called \emph{symmetry basis}, i.e. the basis in which the symmetry in question becomes most evident. We adopt the latter name.

\emph{U(1)-symmetric operator}. 
An operator acting on a physical system with a symmetry can be represented as a block matrix, i.e. as a direct sum of matrices belonging to different symmetry sectors. If restricted to a sub-region of the lattice $\mathcal{L}' \subset \mathcal{L}$ with $L$ number of sites, the symmetry sectors are determined by the different number of allowed particles and the associated degeneracies are determined from the number of ways that particles could be arranged on the sub-region. This can be written as 
\begin{equation}
    \hat{O}_{\mathcal{L}'} = \bigoplus_{n=0}^{L'} \hat{O}_n.
\end{equation}
The operator $\hat{O}_n$ in the $n$-particle sector can be written in the symmetry basis as 
\begin{equation}
\label{Eq:Symmetric op in CD basis}
    \hat{O}_n = \sum_{\mu_n, \mu_n'} \left( O_{n} \right)_{\mu_n,\mu'_n} \ket{n \mu_n}\bra{n \mu'_n}.
\end{equation}
If this extended to the entire lattice $\mathcal{L}$ with the total of $N$ particles, there would be only a single sector, i.e. a single block in the matrix representation of the operator.

The important point is that any observable that is particle-conserving, whether local or global, can be written in a symmetry basis.

\emph{U(1)-symmetric higher tensors}. The block structure of the U(1)-symmetric matrix (or 2-index tensor ) can be extended to higher-order tensors. For example, a 3-index U(1)-symmetric tensor $T$ can be written like
\begin{equation}
    T = \bigoplus_{m_1+m_2=m_3} T_{m_3}^{m_1 m_2},
\end{equation}
where an ``orientation'' has been imposed on the indices of the tensor, where the ``up'' and ``down'' indices imply $m_3 \rightarrow m_1 + m_2$, i.e. an incoming charge $m_3$ splits into two outgoing charges $m_1$ and $m_2$, and consequently, the sum of the two charges $m_1$ and $m_2$ should equal $m_3$. This implies charge conservation. Each block tensor can be written in the symmetry basis as
\begin{align}
        T_{m_3}^{m_1 m_2} = \sum_{\mu_{m_1}, \mu_{m_2}, \mu_{m_3}} &\left( T_{m_3}^{m_1 m_2} \right)_{\mu_{m_2}}^{\mu_{m_1} \mu_{m_2}} \nonumber \\ 
        &  \ket{m_1 \mu_{m_1}; m_2 \mu_{m_2}}\bra{m_3 \mu_{m_3}}.
\end{align}
Again this can either be applied to a sub-region or the whole system. When applied globally to a system with $N$ particles, we can set $m_3=N$ and $m_2= N - m_1$, for some value of $m_1$. Each block tensor is constructed appropriately to encode the amplitudes of the operation to be performed. 

The example can be generalized to any $n$-index tensor, for $n>3$, with the condition that the integers labeling the charge sectors satisfy charge conservation.

\subsection{Two special tensors for bases transforms}
In this section, we introduce two special tensors that would be important later. From now on, we shall restrict our attention to low-order tensors of two and three indices. It is well known that any tensor with a higher order can be brought into lower order either by iteratively decomposing the higher-order tensor into pairs of lower-order tensors or by ``reshaping'' the higher-order tensor into a lower order one while preserving the elements of the tensor. These two approaches have different applications. For our purpose, we follow the latter approach, which can be achieved with two special trivalent tensors called \emph{splitting} and \emph{fusion} tensors, which are introduced below.

\emph{Symmetric Splitting and Fusion tensor}. In fact, the 3-index symmetric tensor $T$ exposed above could have been obtained from a 2-index symmetric matrix by ``splitting'' the outgoing index of the matrix into two new outgoing indices using a special \emph{splitting tensor}. Besides this, there is also a dual object, a \emph{fusion tensor} that can be used to reshape a 3-index tensor into a matrix (i.e. 2-index), by combining the two outgoing indices of $T$ into a new single outgoing index. 

We now focus on the realization of the fusion splitting tensors for U(1) symmetry. On the one hand, let $F$ denote the U(1)-symmetric fusion tensor that encodes the coefficients of a fusion map 
\begin{equation}
    \mathcal{F} : \mathbb{V}_1 \otimes \mathbb{V}_2 \rightarrow \mathbb{V}_{12}.
\end{equation}
The coefficients are set to $1$ for a valid, unique one-to-one map from the product symmetry basis $\{ \ket{n_1 \mu_{n_1}} \otimes \ket{n_2 \mu_{n_2}} \}$ of the two vector spaces $\mathbb{V}_1$ and $\mathbb{V}_2$ to the coupled basis $\{ \ket{n \mu_{n}} \}$ of $\mathbb{V}_{12}$, where $n\equiv n_1+n_2$, otherwise the coefficients are set to $0$. The fusion ($F$) tensor can be written in components as $F_{(n_1,\mu_{n_1})(n_2,\mu_{n_2})}^{(n,\mu_n)}$. To make this concrete, consider the example of coarse-graining two fermionic sites into a single composite site. In terms of the symmetry basis, the basis of a single site is $\{ \ket{0,1}, \ket{1,1} \}$, where the first entry in $\ket{\cdot, \cdot}$ is the charge $\{0, 1\}$ and the second is the degeneracy index of the charges (which is just $1$ in this case for both charges). The basis of the two-site composite is $\{\ket{0,1}, \ket{1,1}, \ket{1,2}, \ket{2,1} \}$. The obtained charges and degeneracies are derived from the fusion of the charges on the two sites, which is based on the addition ``fusion rules'':
\begin{equation}
    0 \times 0 \rightarrow 0, ~ 0 \times 1 = 1 \times 0 \rightarrow 1, ~ 1 \times 1 \rightarrow 2.
\end{equation}
The elements of the fusion tensor $F$ can be taken as
\begin{equation}
    F_{(0,1)(0,1)}^{(0,1)} = F_{(0,1)(1,1)}^{(1,1)} = F_{(1,1)(0,1)}^{(1,2)} = F_{(1,1)(1,1)}^{(2,1)} = 1,
\end{equation}
and other elements are set to zero.

On the other hand, the dual splitting map $\mathcal{S}:\mathbb{V}_{12}  \rightarrow  \mathbb{V}_1 \otimes \mathbb{V}_2$ and its corresponding tensor $S$ can be constructed easily by reversing the direction of the arrow of the fusion map. The tensor $S$ can be written in components as $S_{(n,\mu_n)}^{(n_1, \mu_{n_1}) (n_2, \mu_{n_2})}$. The non-zero elements are given as
\begin{equation}
    S^{(0,1)(0,1)}_{(0,1)} = S^{(0,1)(1,1)}_{(1,1)} = S^{(1,1)(0,1)}_{(1,2)} = S^{(1,1)(1,1)}_{(2,1)} = 1,
\end{equation}
and zero otherwise. 

It is important to realize that the splitting and fusion tensors should be constructed such that they satisfy the following identities:
\begin{equation}
    \begin{matrix}
        \includegraphics[width=0.8\columnwidth]{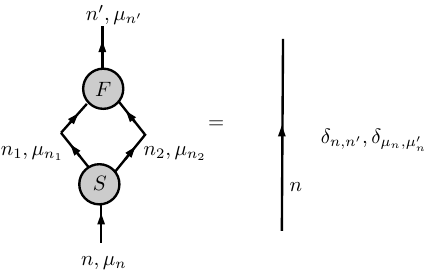}
    \end{matrix}
\end{equation}
and 
\begin{equation}
    \begin{matrix}
        \includegraphics[width=0.8\columnwidth]{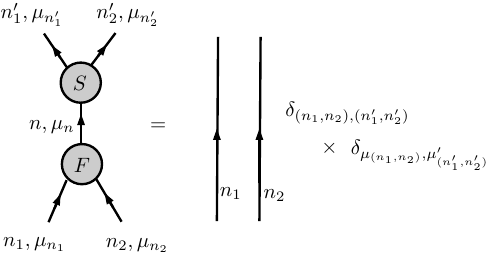}
    \end{matrix}
\end{equation}
We verified that this is true for our choices above.

\subsection{Elementary form of two-site symmetric operators}
In our discussion above, we proposed to construct higher-order tensors by composing lower-order tensors using appropriate fusion and splitting tensors. We now show a specific example: a symmetric $4$-index tensor can be  ``composed'' from a  network of ``smaller'' tensors as 
\begin{equation}
\label{Eq:Brick tensor decomposed}
    \begin{matrix}
        \includegraphics[width=0.8\columnwidth]{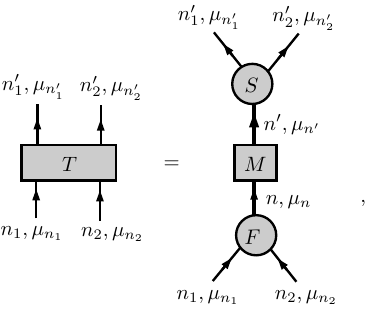}
    \end{matrix}
\end{equation}
where we first apply a fusion tensor to couple the two sites into one, transforming the product symmetry basis into a coupled one. The matrix $M$ is written in this coupled basis, and finally, the splitting tensor is used to transform back into the original product basis. For U(1) symmetry, charge conservation implies $n_1 + n_2 = n = n' = n'_1 + n'_2$. In the above diagram, the symmetric matrix $M$, which is the ``reshaped version'' of tensor $T$ is denoted with fattened lines. The matrix has a  block structure, $M = \oplus_n M_n$. 

For the example of two fermionic sites, the symmetric matrix $M$ can be written as
\begin{equation}
    M_0 \oplus M_1 \oplus M_2,
\end{equation}
where the matrices corresponding to the different charge sectors with charges $0,1,2$ are
\begin{equation}
\label{Eq:2Sites charge sectors matrix}
    M_0 = (a), \quad 
    M_1 = 
    \begin{pmatrix}
        b & c \\
        d & e
    \end{pmatrix}
    \quad 
    M_2 = (f),
\end{equation}
for $a,b,c,d,e,f \in \mathbb{C}$. The sizes of the matrices are specified by their degeneracies. 

The ``elementary form'' for any two-site symmetric operator shown in Eq.\eqref{Eq:Brick tensor decomposed} will be used as our ``template''  for constructing particle-conserving symmetric quantum gates, which are subsequently used to construct particle-conserving symmetric quantum circuits. These are the focus of the next two sections.

\section{Construction of symmetric quantum gates}
\label{Sec:Symmetric circuit elements}
The symmetric tensor decomposition presented in the previous section consists of three types of tensor objects, namely, a fusion and splitting 3-index tensor, and a charge-conserving matrix (or 2-index tensor). Our approach to building symmetric quantum circuits is to show how to construct the quantum gates realizing these symmetric tensor objects, with the only condition that the gates are unitary. Similar to a symmetric tensor network, we shall view a symmetric quantum circuit as a network of symmetric quantum gates that are composed together.

In this section, we shall lay out a theoretical principle of how to construct elementary quantum gates of symmetric operators from a bottom-to-top approach and then specialize to the application of constructing symmetric quantum gates on two fermionic sites. In the next section, we will use the symmetric quantum gates to construct a ``brick-wall'' particle-conserving variational quantum circuit.

\subsection{The general idea}
As every quantum circuit can be built with one- and two-qubit quantum gates,\cite{nielsen2002quantum} we, therefore, focus on two-qubit gates, outlining the steps involved in constructing symmetric operators. (The case of one-qubit symmetric will become obvious).

Let $\mathcal{U}_{\mathrm{sym}}: \mathbb{V^{(A)}} \otimes \mathbb{V^{(B)}} \rightarrow \mathbb{V^{(A)}} \otimes \mathbb{V^{(B)}}$ be a symmetric unitary operator on two qubits. Similar to the right-hand side decomposition of Eq.~\eqref{Eq:Brick tensor decomposed}, we follow the steps below to construct $\mathcal{U}_{\mathrm{sym}}$:

\begin{enumerate}
    \item First, apply a ``fusion'' map $\mathcal{F}$ to transform from the product space $\mathbb{V^{(A)}} \otimes \mathbb{V^{(B)}}$ to a composite space $\mathbb{V^{(AB)}}$ as $\mathcal{F} : \mathbb{V^{(A)}} \otimes \mathbb{V^{(B)}} \rightarrow \mathbb{V^{(AB)}}$.

    \item Then construct a charge-conserving operator $\mathcal{Q}$ on the composite space $\mathbb{V^{(AB)}}$ in the symmetry basis.

    \item Finally, transform back into the product space $\mathbb{V^{(A)}} \otimes \mathbb{V^{(B)}}$ using a ``splitting'' map $\mathcal{S} : \mathbb{V^{(AB)}} \rightarrow \mathbb{V^{(A)}} \otimes \mathbb{V^{(B)}}$.
\end{enumerate}
Composing all the operators together gives
\begin{equation}
    \mathcal{U}_{\mathrm{sym}} = \mathcal{S} \circ \mathcal{Q} \circ \mathcal{F},
\end{equation}
which can be summarized with the commutative diagram 
\begin{equation}
\begin{matrix}
\begin{tikzcd}[sep=huge]
\mathbb{V}^{(A)}  \otimes \mathbb{V}^{(B)}  \arrow[r,shift left,"\mathcal{U}_{\mathrm{sym}}"]  \arrow[d,shift left, "\mathcal{F}"]  & \mathbb{V}^{(A)}  \otimes \mathbb{V}^{(B)} \\
\mathbb{V}^{(AB)} \arrow[r,shift left,"\mathcal{Q}"] & \mathbb{V}^{(AB)} \arrow[u,shift left,"\mathcal{S}"]
\end{tikzcd}
\end{matrix}.
\end{equation}

The exact matrix representations of the fusion and splitting maps will depend on the symmetry group in question, so we defer it to a later section, where we present the specific case of particle conservation. Although, we now present the construction of symmetric gates in the symmetry basis of a composite space $\mathbb{V^{(AB)}}$.

\subsection{Symmetric operator quantum circuit in the symmetry basis}
\label{Sec:Symmetric Op quantum circuit}
In Sec. \ref{Sec:U1Tensors} we showed that symmetric operators can be written as charge-conserving 2-index tensors over the symmetry basis. In other words, as block matrices, where each block matrix is associated with each conserved charge. 

To construct quantum circuits realizing symmetric operators, we shall make use of the following proposition, which is only valid in the symmetry basis.

\begin{theorem}
Charge-conserving operators can be realized as controlled operations, and vice versa.
\label{Thm: Charge-conserving as controlled operators}
\end{theorem}

We will ``prove'' this statement in two sections. In the first, we deal with the first part of the statement, that \emph{charge-conserving operators can be implemented with controlled operations}. In the second, we will show that \emph{controlled operations can be seen as charge-conserving operators}, in the symmetry basis.

\subsubsection{Charge-conserving operators as controlled operations}
Let $\hat{Q}$ be a symmetric operator (i.e. charge-conserving). In the symmetry basis, we express it as
\begin{equation}
\hat{Q} = \sum_{a,i_a, i'_a} \left( Q_a  \right)_{i_a, i'_a} \ket{a,i_a} \bra{a,i'_a},
\end{equation}
where $a$ labels the conserved charges and the indices $i_a, i'_a$ enumerate the degeneracy of each charge $a$. This can be re-organized as
\begin{equation}
\hat{Q} = \sum_a \ket{a}\bra{a} \otimes \left(\sum_{i_a,i'_a} \left(Q_a\right)_{i_a,i'_a} \ket{i_a} \bra{i'_a}  \right).
\end{equation}
The expression in the big brackets can be written succinctly as $\hat{Q}^{(a)}$: the operator acting over the degeneracy space of charge $a$. Therefore,
\begin{equation}
\hat{Q} = \sum_{a} \ket{a}\bra{a} \otimes \hat{Q}^{(a)}.
\end{equation}
This is nothing other than a controlled operation, where the operator $\hat{Q}^{(a)}$---which has to be unitary---is applied if the charge is determined to be $a$. We shall use two sets of qubits to encode this operator, namely (charge) control qubits and (degeneracy) target qubits. The $\ket{a}\bra{a}$ is a projector that acts on the (charge) control qubits, and it conserves charge, while $\hat{Q}^{(a)}$ is the corresponding degeneracy operator that acts on the (degeneracy) target qubits. The above equation says that if the (charge) control qubits are in a state with charge $a$, then apply operator $\hat{Q}^{(a)}$ on the (degeneracy) target qubits.

Rather than giving a generic circuit diagram for symmetric operators in the symmetry basis, we present a simple example for two qubits, without loss of generality. Assume there is one control and one target qubit, the operator becomes
\begin{equation}
\label{Eq:two-qubit symmetric op}
    \hat{Q} = \ket{0}\bra{0} \otimes \hat{Q}^{(0)} + \ket{1}\bra{1} \otimes \hat{Q}^{(1)}. 
\end{equation}
The circuit for $\hat{Q}$ is 
\begin{equation}
\label{Eq:Symmetric operator circuit}
\begin{matrix}
    \Qcircuit @C=1em @R=.7em {
 & \multigate{1}{\hat{Q}} &  \qw \\
 & \ghost{\hat{Q}}        &   \qw  
}
\end{matrix}
=
\begin{matrix}
 \Qcircuit @C=1.5em @R=1.5em {
 & \ctrlo{1}     & \ctrl{1}       & \qw \\
 & \gate{\hat{Q}^{(0)}} & \gate{\hat{Q}^{(1)}}  & \qw
}
\end{matrix}.
\end{equation}
 
This example can be generalized by using more qubits to either encode more charge sectors and/or increase the degeneracy of the charge sectors. Therefore, charge-conserving operators can be realized as controlled operations, with the constraint that the degeneracy operators are unitary.

The one-qubit case is trivial. The $\hat{Q}^{(0)}$ and $\hat{Q}^{(1)}$ in Eq.~\eqref{Eq:two-qubit symmetric op} can only be complex phases, and hence $\hat{Q}$ is a diagonal $2\times2$ matrix.

\subsubsection{Controlled operations are charge-conserving operators}
We now ``prove'' the reverse statement, that \emph{controlled operations can be represented as charge-conserving operators}.

This can be seen easily if we group together all the control qubits to encode the conserved charges, and the target qubits to encode the degeneracies of the conserved charges. We can represent the charge degrees of freedom of the control qubits with projectors $\ket{a} \bra{a}$ for every charge $a$ in the set of conserved charges. The corresponding unitary operator $\hat{Q}^{(a)}$ is applied on the (degeneracy) target qubits, which  also depends on the charge $a$ of the control qubit. The whole operation can be represented as
\begin{equation}
    \hat{Q} = \sum_{a} \ket{a}\bra{a} \otimes \hat{Q}^{(a)}.
\end{equation}
Therefore, controlled operations can realize charge-conserving symmetric operators, if working in a symmetry basis.

\subsection{Application to two fermionic sites}
\label{Sec:ApplyTwoSites}
We now apply the general results of the previous two sections to the example of two fermionic sites. We use them to construct the symmetric quantum gates that would be used later to build particle-conserving variational quantum circuits.

\subsubsection{Basis mapping}
We start with a discussion of the basis mapping. For a single site, the basis of a single qubit can  be mapped to the occupation number basis as $\ket{0} \leftrightarrow \ket{n=0}$ and $\ket{1} \leftrightarrow \ket{n=1}$. For two sites, the basis is (ordered as): $$\ket{00}, ~ \ket{01}, ~ \ket{10}, ~ \ket{11}.$$ In terms of number symmetry basis $\{ \ket{n,\mu_n} \}$, using the addition fusion rule, the basis are:
\begin{align}
    &\ket{00} \rightarrow \ket{0,1} \nonumber, \quad 
    \ket{01} \rightarrow \ket{1,1} \nonumber \\
    &\ket{10} \rightarrow \ket{1,2} \nonumber,  \quad
    \ket{11} \rightarrow \ket{2,1},
\end{align}
where the first entry is the charge and the second is the degeneracy index. A generic block matrix in the symmetry basis would be
\begin{equation}
\label{Eq:Op M in U1 basis}
M = 
    \begin{pmatrix}
    a &   &    & \\
      & b & c  & \\
      & d & e  & \\
      &   &  & f
    \end{pmatrix},
\end{equation}
respectively for charge sectors $0,1$, and $2$,  where $a,b,c,d,e,f \in \mathbb{C}$. Because there are three different charges: $0,1,2$, it seems that more than a single qubit would be needed to encode the charges on two sites. But if particle number parity conservation is used instead, a single qubit can still be used for the encoding, where the fusion rule is addition modulo $2$ for $\mathbb{Z}_2$ group. Under this fusion rule, the basis states are:
\begin{align}
    &\ket{00} \rightarrow \ket{0,1} \nonumber, \quad 
    \ket{01} \rightarrow \ket{1,1} \nonumber \\
    &\ket{10} \rightarrow \ket{1,2} ,  \quad
    \ket{11} \rightarrow \ket{0,2} ,
\end{align}
where there are now only two charges: $0,1$, on the two sites. The symmetric block matrix for this new basis is
\begin{equation}
\label{Eq:Op M with Z2}
M =  M_{0} \oplus M_1 . 
\end{equation}
To maintain the same number of non-zero elements as in Eq.\eqref{Eq:Op M in U1 basis}, we constrain the $M_0$ to be diagonal, so that $M$ is
\begin{equation}
M = 
 \begin{pmatrix}
        a & 0  \\
        0 & f
    \end{pmatrix}
    \oplus 
    \begin{pmatrix}
        b & c \\
        d & e
    \end{pmatrix},
\end{equation}
for $a,b,c,d,e,f \in \mathbb{C}$. To use the symmetric operator as a quantum gate, the block matrices are constrained to be unitary. The circuit diagram for the symmetric operator $M$ in Eq.\eqref{Eq:Op M with Z2} will have the form shown in Eq.~\eqref{Eq:Symmetric operator circuit}.

\subsubsection{Parameterizations of symmetric two-qubit gates in the symmetry basis}
In this section, we show two different physically-motivated ways of constructing two-qubit symmetric  gates. Unless otherwise stated, we work in a symmetry basis.

\emph{1. Charge-1 sector parameterization}: The first approach we consider is to find a ``generating'' unitary operator for the generating charge, $1$---since it can be used to obtain all other charges; $0$ is the identity charge. The generating unitary can be used to create a quantum state of a single particle on two fermionic sites. This approach is similar to a method in Ref.~\cite{gard2020efficient}. 

A generic state for a system with one particle on two sites can be written in the symmetry basis as 
\begin{equation}
\ket{\Psi} = \alpha \ket{1,1} + \beta \ket{1,2}, \quad \alpha, \beta \in \mathbb{C},
\end{equation}
which can be rewritten as 
\begin{equation}
\ket{\Psi} = \ket{1} \otimes \left(\alpha \ket{1} + \beta \ket{2} \right),
\end{equation}
where the first term is the ``charge state'' and the second term is the ``degeneracy state.'' If we impose normalization and ignore a global phase, only two real parameters are needed to specify $\ket{\Psi}$. Nothing stops us from choosing this unitary as 
\begin{equation}
V = 
\begin{pmatrix}
 - \mathrm{sin} \theta & e^{-i\phi} \mathrm{cos} \theta \\
e^{i\phi} \mathrm{cos} \theta  & \mathrm{sin} \theta\\
\end{pmatrix},
\end{equation}
having the two real parameters needed to parameterize the quantum state. It can be checked that $V$ can be written as
\begin{equation}
    V = R_z(\phi) R_y(2\theta) R_z(\phi) X, 
    \label{Eq:DecompositionV}
\end{equation}
where $X$ is the Pauli $\sigma_x$, and $R_x(\theta) = e^{-i \frac{\theta}{2} X}$, and similar expressions for $R_y$ and $R_z$ in terms of their Pauli matrices $\sigma_y$ and $\sigma_z$. If we set $U = R_z(\phi) R_y(\theta)$, $V$ can be rewritten as
\begin{equation}
    V = U X U^{\dagger}.
\end{equation}
It is easy to check that this gives Eq.~\eqref{Eq:DecompositionV}, if we use the identities \begin{equation}
    X R_y(\theta) X = R_y(-\theta), \quad X R_z(\theta) X = R_z(-\theta). 
\end{equation}
(The matrix representations of these operators and their identities can be recalled from Ref.~\cite{nielsen2002quantum}.) To construct the symmetric operator over the two sites, we do a controlled operation on $V$. A controlled operation of $V$ has the usual interpretation: apply the unitary operator $V$ if the charge qubit is set to $1$, otherwise do nothing (i.e. apply the identity). The circuit diagram is 
\begin{equation}
\mathcal{C}_1(V)  = 
    \begin{matrix}
        \Qcircuit @C=0.6em @R=1em {
            & \ctrl{1} & \qw \\
            & \gate{V(\theta, \phi)} & \qw
        }        
    \end{matrix} = 
    \begin{matrix}
        \Qcircuit @C=0.6em @R=1em {
            & \qw               &  \ctrl{1} & \qw & \qw \\
            & \gate{U^{\dagger}(\theta, \phi)} & \targ     & \gate{U(\theta, \phi)} & \qw 
        }        
    \end{matrix} ,
\end{equation}
where the circuit decomposition for $\mathcal{C}_1(V)$ involves only $1$ CNOT gate and two single qubit unitaries. The circuit has the action:
\begin{align}
&\ket{0,1} \rightarrow  \ket{0,1} \\
&\ket{0, 2} \rightarrow  \ket{0, 2} \\
&\ket{1,1} \rightarrow  - \mathrm{sin} \theta \ket{1,1} + e^{i \phi} \mathrm{cos} \theta \ket{1,2}\\
&\ket{1,2} \rightarrow  e^{-i \phi}\mathrm{cos}\theta \ket{1,1} + \mathrm{sin} \theta \ket{1,2}.
\end{align}
Written explicitly, the matrix representation of the controlled operation $\mathcal{C}_1(V)$ is
\begin{equation}
\left[\mathcal{C}_1(V) \right]  = 
\begin{pmatrix}
1 & 0 & 0 & 0 \\
0 & 1 & 0 & 0 \\
0 & 0 & -\mathrm{sin} \theta & e^{ - i\phi} \mathrm{cos} \theta \\
0 & 0 &  e^{i \phi}\mathrm{cos}\theta & \mathrm{sin} \theta
\end{pmatrix},
\end{equation}
which is just $\mathbb{I}_2 \oplus V$. Therefore, with respect to Eq.~\eqref{Eq:Op M with Z2}, the unitary needed for each charge sector is $M_0 = I_2$ and $M_1 = V$.

\emph{2. A model-inspired parameterization}: Our second approach is to use inspiration from the problem considered. A typical model in our case is a tight-binding model in 1D. We use the nature of the fermionic local Hamiltonians as a guide in building a basic parameterized quantum gate. On two sites, the most basic energetic processes are either hopping or particle-particle interaction (assuming no external on-site potential). The local Hamiltonian on two sites can be written as 
\begin{equation}
    \hat{h} = -t \left(\ket{10}\bra{01} + \ket{01} \bra{10} \right) -\lambda \hat{n}_1 \hat{n}_2,
\end{equation}
where $t \in \mathbb{R}$ is tunneling amplitude and $\lambda$ is interaction strength. The time evolution of a quantum system can be approximated using Trotter decomposition, where the operator $\hat{u} = e^{-i\hat{h}\delta \tau}$ are applied locally on pairs of qubits alternately for some specified time $T$. We can again write $\hat{h}$ in the symmetry basis as
\begin{align}
    \hat{h} & = \hat{h}_0 \oplus \hat{h}_1 \nonumber \\
            & = 
            \begin{pmatrix}
                0 & 0 \\
                0 & -\lambda
            \end{pmatrix}
            \oplus
            \begin{pmatrix}
                0  & -t \\
                -t &  0 
            \end{pmatrix}, \\
        & =  -\lambda \hat{n}  - t \hat{X}. \nonumber
\end{align}
The local time evolution is 
\begin{align}
    \hat{u} & = \hat{u}_0 \oplus \hat{u}_1 \nonumber \\
            & = e^{i \phi \hat{n}}  \oplus e^{i \frac{\theta}{2}\hat{X}}  \\
            & = P(\phi) \oplus R_x^{\dagger}(\theta),
\end{align}
where $\phi = \lambda \delta \tau$ and $\theta = 2 t \delta \tau$. The $P(\phi)$ and $R_x(\theta)$
are the matrices
$$P(\phi) = \begin{pmatrix}
    1 & 0 \\
    0 & e^{i\phi}
\end{pmatrix}$$
and 
$$
R_x(\theta) = e^{-i \frac{\theta}{2} X} = \begin{pmatrix}
\cos{\theta/2}    & -i\sin{\theta/2}   \\
-i \sin{\theta/2} &  \cos{\theta/2}
\end{pmatrix}.
$$
The circuit diagram for $\hat{u}$ is 
\begin{equation}
\hat{u}(\theta, \phi) = 
   \begin{matrix}
        \Qcircuit @C=1.2em @R=1.5em {
            & \ctrlo{1} & \ctrl{1} & \qw \\
            & \gate{P(\phi)} &\gate{R_x^{\dagger}(\theta)} & \qw
        }        
    \end{matrix} .
\end{equation}
{We will use $\hat{u}(\theta, \phi)$ as a charge-conserving two-qubit parameterized gate, where $\theta, \phi$ are parameters.} The circuit can be understood as ``apply $P(\phi)$ on the (degeneracy) target qubit if the (charge) control qubit is $0$ and apply the rotation gate $R^{\dagger}_x(\theta)$ on (degeneracy) target qubit if the (charge) control qubit is $1$.'' 

The symmetric operators constructed with both methods above are in the symmetry basis. As explained in Sec.~\ref{Sec:U1Tensors}, we will need the fusion and splitting tensors to transform back and forth between the product basis and the symmetry basis. As the symmetric operators are constructed with an underlying $\mathbb{Z}_2$ symmetry, the quantum gates realizing the fusion and splitting tensors will also respect the $\mathbb{Z}_2$ fusion rule.

\subsubsection{Quantum gates of $\mathbb{Z}_2$ fusion and splitting maps}
To complete building the symmetric quantum gate for two fermionic sites, we have to find the quantum gates realizing $\mathbb{Z}_2$ fusion and splitting tensors. The requirement on the ``fusion tensor'' gate is that it should map from the product basis of two sites to the symmetry basis. And the ``splitting tensor'' gate should do the reverse. In other words, the fusion map is to ``enter'' the symmetry basis, and the splitting map is to ``step out'' back to the original basis. 

The choice of the fusion map that we use is: 
\begin{align}
& \ket{00} \rightarrow \ket{0,1},  \quad \ket{01} \rightarrow \ket{1,2},  \nonumber \\
& \ket{10} \rightarrow \ket{1,1}, \quad \ket{11} \rightarrow \ket{0,2} \nonumber.
\end{align}
If the ordering of the product basis is $\{\ket{00}, \ket{01}, \ket{10}, \ket{11}\}$ and that of $\mathbb{Z}_2$-symmetric basis is $\{\ket{0,1}, \ket{0,2}, \ket{1,1}, \ket{1,2}\}$, the ``fusion tensor'' gate can be written as the matrix
\begin{equation}
F ~=   
    \begin{pmatrix}
    1 & 0 & 0 & 0 \\
    0 & 0 & 0 & 1 \\
    0 & 0 & 1 & 0 \\
    0 & 1 & 0 & 0
    \end{pmatrix}.
\end{equation}
This matrix corresponds to the ``reflected version'' of the CNOT gate, where the bottom qubit is the control and the top qubit is the target,
\begin{equation}
\label{Eq:FusionGate}
F = 
    \begin{matrix}
        \Qcircuit @C=1em @R=1em {
        & \targ & \qw \\
        & \ctrl{-1} & \qw
        }    
    \end{matrix} \quad .
\end{equation}

Another possible $\mathbb{Z}_2$ fusion map that could have been used is 
\begin{align}
& \ket{00} \rightarrow \ket{0,1},  \quad \ket{01} \rightarrow \ket{1,1},  \nonumber \\
& \ket{10} \rightarrow \ket{1,2}, \quad \ket{11} \rightarrow \ket{0,2} \nonumber,
\end{align}
which corresponds to the matrix
\begin{equation}
\begin{pmatrix}
1 & 0 & 0 & 0 \\
0 & 0 & 0 & 1 \\
0 & 1 & 0 & 0 \\
0 & 0 & 1 & 0 
\end{pmatrix}.
\end{equation}
But this is realized as double CNOT,
\begin{equation}
    \begin{matrix}
        \Qcircuit @C=1em @R=1em {
        & \targ & \ctrl{1} & \qw & \\
        & \ctrl{-1} & \targ & \qw
        }    
    \end{matrix}.
\end{equation}

We will use the first option, in Eq.~\eqref{Eq:FusionGate}, as it minimizes the number of CNOT operations. 

The ``splitting tensor'' gate $S$ can be obtained easily from the ``fusion tensor'' gate $F$ by inverting its matrix representation. Note that $F$ is an involution, i.e. $F^2 = \mathbb{I}_4$, so $F = F^{-1}$, and therefore the matrix for $F$ and $S$ are the same.

\subsubsection{Particle-conserving quantum gate module}
In Sec.~\ref{Sec:U1Tensors} we showed that a symmetric $4$-index tensor may be obtained by composing some three tensors, namely a fusion tensor $F$, a charge-conserving matrix $M$, and a splitting tensor $S$. Having obtained all the quantum gates corresponding to these tensors, we can similarly compose them to build a symmetric quantum gate that locally conserves particle number. We refer to the local particle-conserving exchange gates as a ``module.''

In building the local symmetric gate module, we start with the ``fusion gate,'' which brings the product basis on two sites into the symmetry basis, and then compose with the charge-conserving gate, and finally apply the ``splitting gate'' to transform back to the product basis. Corresponding to the two symmetric gates presented above in the symmetry basis, the two different gate modules, denoted by $A$ and $B$, are given below as

\begin{equation}
\label{Eq:A gate}
\begin{matrix}
    \Qcircuit @C=0.4em @R=.7em {
 & \multigate{1}{A(\theta, \phi)} &  \qw \\
 & \ghost{A(\theta,\phi)}        &   \qw  
}
\end{matrix} = 
   \begin{matrix}
        \Qcircuit @C=0.4em @R=1.5em {
            & \targ        &  \qw                           &  \ctrl{1} & \qw           & \qw           &  \targ     & \qw  \\
            & \ctrl{-1}    &  \gate{U^{\dagger}(\theta, \phi)}   & \targ     & \gate{U(\theta, \phi)}   & \qw            & \ctrl{-1}. & \qw
        }        
    \end{matrix} \quad,
\end{equation}
and 
\begin{equation}
\begin{matrix}
    \Qcircuit @C=0.5em @R=.7em {
 & \multigate{1}{B(\theta, \phi)} &  \qw \\
 & \ghost{B(\theta,\phi)}        &   \qw  
}
\end{matrix} = 
   \begin{matrix}
        \Qcircuit @C=0.5em @R=1.5em {
            & \targ        &  \ctrlo{1}                   &  \ctrl{1}           & \qw           &  \targ     & \qw  \\
        & \ctrl{-1}    &  \gate{P(\phi)}          &   \gate{R_x^{\dagger}(\theta)}   & \qw            & \ctrl{-1}. & \qw
        }        
    \end{matrix} \quad.
\end{equation}
 
The matrices corresponding to these gates can be obtained easily to be
\begin{equation}
\label{gate A}
    A(\theta, \phi) = 
    \begin{pmatrix}
        1 & 0 & 0 & 0 \\
        0 & \sin{\theta} & e^{i\phi} \cos{\theta} & 0 \\
        0 & e^{-i\phi}\cos{\theta} & -\sin{\theta} & 0 \\
        0 & 0 & 0 & 1
    \end{pmatrix},
\end{equation}
and
\begin{equation}
\label{gate B}
    B(\theta, \phi) = 
    \begin{pmatrix}
        1 & 0 & 0 & 0 \\
        0 & \cos{\frac{\theta}{2}} & i \sin{\frac{\theta}{2}} & 0 \\
        0 & i \sin{\frac{\theta}{2}} & \cos{\frac{\theta}{2}} & 0 \\
        0 & 0 & 0 & e^{i \phi}
    \end{pmatrix}.
\end{equation}
{A couple of notes are in order. First, we checked that the two gates are not locally equivalent. Hence, we can use them independently to build particle-conserving variational quantum circuits. Secondly, we note that the $A(\theta, \phi)$ gate obtained is the same as the one used in Refs.~\cite{barkoutsos2018quantum,gard2020efficient}, if we let $\theta \rightarrow \frac{\pi}{2} - \theta$. But note that we obtained this gate directly by composing elementary gates. Our method can be viewed as a bottom-to-top approach, which is different from a top-to-bottom approach, where a two-qubit general unitary gate is specialized to be particle-conserving, and then  decomposed into its elementary gates. Thirdly, we also note that both the $A(\theta, \phi)$ and $B(\theta, \phi)$ gates are similar to the particle-conserving gates that are used to build hardware-efficient ansatzes for quantum chemistry (see Ref.~\cite{barkoutsos2018quantum}).} 

The two gates above are similar to examples of particle-conserving exchange-type gates that have been previously to construct particle-conserving quantum circuits. Note that the two gates were derived using the two parameterizations discussed above. If other types of parameterizations are considered, more symmetric gates can possibly be derived.

In the next section, we use the two gates as basic ``blocks'' to build particle-conserving quantum circuits. While we can in principle mix both gates in a single composition, we will only build ``homogenous'' symmetric quantum circuits with either gate $A$ or $B$. For ease of notation, the circuits will be called $C_A$ and $C_B$ respectively.

\section{Particle-Conserving Variational Quantum Circuits}
\label{Sec:Number Conserving Circuit}
We now consider how to construct VQCs that globally conserve the total number of particles, $N$ on a lattice of $L$ sites.

\subsection{Particle-conserving quantum states}
The Fock space $\mathcal{H}_{N,L}$ for a system of $N$ spinless electrons on a lattice of $L$ sites is spanned by the occupation number basis states $\{ \ket{n_1, n_2,\ldots, n_L}\}$,
such that $\sum_{i=1}^L n_i = N$, where $N$ sites are occupied and the rest are unoccupied. The dimension of $\mathcal{H}_{L,N}$ is $d_{N,L} = \left( \begin{matrix} L \\ N \end{matrix} \right)$. 

A general pure quantum state can be written in the occupation number basis as 
\begin{equation}
\ket{\Psi_{N,L}} = \sum_{n_1, n_2,  \ldots, n_L} c_{n_1, n_2,  \ldots, n_L} \ket{n_1, n_2,  \ldots, n_L}, 
\end{equation}
with $d_N$ number of complex coefficients $c_{n_1, \ldots, n_L} \in \mathbb{C}$, and $n_i \in \{ 0, 1 \}$. By applying the constraint of normalization and neglecting a global factor, the number of coefficients can be reduced to $2(d_{N,L} -1) $ real numbers. An ansatz for $\ket{\Psi_{N,L}}$ therefore needs that number of real parameters. Such a variational ansatz can be created on a quantum processor by applying a network of one-qubit and two-qubit parameterized gates on an initial state $\ket{\psi_0}$, such that the number of variational parameters equals the number needed to span the subspace.

\subsection{U(1)-symmetric quantum circuit}
\label{Sec:U1 symmetric quantum circuit}
As $\ket{\psi_{N,L}}$ is a U(1)-symmetric state with a global conserved particle number $N$, one approach to building a particle-conserving quantum circuit is to compose symmetric quantum gates that locally conserve particle number. Having shown how to construct such local gates in Sec.~\ref{Sec:Symmetric circuit elements}, namely $A(\theta, \phi)$ and $B(\theta, \phi)$ two-qubit gates, we now use them to build the variational quantum circuits. 

Let $\hat{U}(\pmb{\theta, \phi})$ be a particle-conserving unitary gate for the entire system, where $\pmb{\theta, \phi}$ are the sets of all parameters, $\theta,\phi$. Therefore the parameterized state is 
\begin{equation}
\ket{\Psi(\pmb{\theta}, \pmb{\phi})} = \hat{U}(\pmb{\theta},\pmb{\phi}) \ket{\psi_0},
\end{equation}
where $\ket{\psi_0}$ is an initial state within the relevant subspace $\mathcal{H}_{N,L}$. We create this initial state $\ket{\psi_0}$ by setting $N$ qubits to the logical state $\ket{1}$ and the remaining $(L-N)$ to the logical state $\ket{0}$.  

An important question is how many basic symmetric gates ($A$ or $B$) are needed to build the circuit for $\hat{U}(\pmb{\theta},\pmb{\phi}) $. Since each basic symmetric gate contributes two real parameters, then $(d_{N,L} - 1)$ number of basic symmetric gates should in principle do. However, it was found in Ref.~\cite{gard2020efficient} that a direct composition of this number of basic gates do not give a quantum circuit that spans the relevant symmetric subspace. Rather they found numerically, using a fidelity test, that a total of $d_{N,L}$ number of basic gates are needed. But doing this brings the total number of parameters to $2d_{N,L}$, and only  by fixing two of the parameters was the  total number reduced to $2(d_{N,L}-1)$---the actual number required to span $\mathcal{H}_{N,L}$. Following this result, we present an algorithm that can be used to build particle-conserving variational quantum circuits. Let the circuits be denoted $C_A(\pmb{\theta}, \pmb{\phi})$ and $C_B(\pmb{\theta}, \pmb{\phi})$, which are composed entirely with $A(\theta, \phi)$ and $B(\theta, \phi)$ respectively. 

\begin{figure}
\[ \Qcircuit @C=1em @R=1em {
\ket{0} &	&	\qw	&	\multigate{1}{A}	&	\qw	&	\qw	&		&		&		&	\qw	&	\multigate{1}{A}	&	\qw	&	\qw	\\
\ket{0} &	&	\gate{X}	&	\ghost{A}	&	\multigate{1}{A}	&	\qw	&		&		&		&	\qw	&	\ghost{A}	&	\multigate{1}{A}	&	\qw	\\
\ket{0} &	&	\qw	&	\qw	&	\ghost{A}	&	\qw	&		&		&		&	\qw	&	\qw	&	\ghost{A}	&	\qw	\\
	&		&		&		&		&		&		&		&		&		&		&		\\
\vdots	&		&		&	\vdots	&		&		&	\ddots	&		&		&		&		&	\vdots	\\
	&		&		&		&		&		&		&		&		&		&		&		\\
\ket{0} &	&	\gate{X}	&	\qw	&	\multigate{1}{A}	&	\qw	&		&		&		&	\qw	&	\qw	&	\multigate{1}{A}	&	\qw	\\
\ket{0} &	&	\qw	&	\multigate{1}{A}	&	\ghost{A}	&	\qw	&		&		&		&	\qw	&	\multigate{1}{A}	&	\ghost{A}	&	\qw	\\
\ket{0} &	&	\gate{X}	&	\ghost{A}	&	\qw	&	\qw	&		&		&		&	\qw	&	\ghost{A}	&	\qw	&	\qw
} \]
\caption{The general circuit construction for generating an ansatz state for a system of $N$ electrons on $L$ sites. It consists of $N$ initial X gates to bring the quantum processor into the $\mathcal{H}_{N,L}$ Fock space (as explained in the text), and then followed by a sequence of $A(\theta,\phi)$ gate composed in an alternate pattern on odd and even sites. Even though it is not shown in the diagram, note that each A gate has different parameters, except the two parameters that are fixed, as explained in the text.}
\label{Fig:QuantumCircuit_A}
\end{figure}
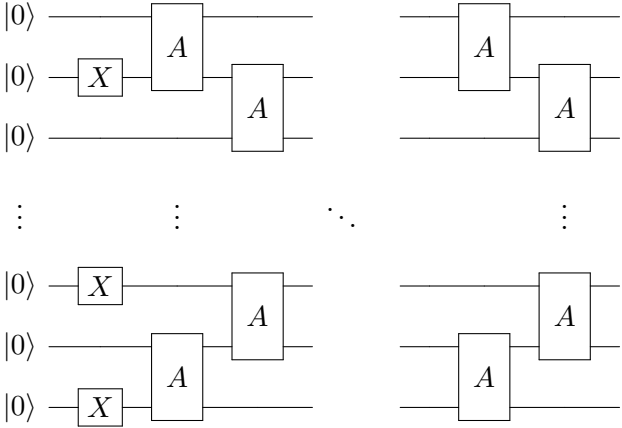

The circuit diagram that generates $C_A(\pmb{\theta}, \pmb{\phi})$ is shown in Fig.~\ref{Fig:QuantumCircuit_A}. (The construction for $C_B(\pmb{\theta}, \pmb{\phi})$ is similar but using the $B(\theta, \phi)$ gate.) The algorithm behind the construction of this circuit is presented below:
\begin{itemize}
\item Apply X gates to N number of qubits to bring the quantum processor into the N-particle subspace.
\item Apply a first layer of A-gate (B-gate) on odd sites, followed by another layer on even sites. 
\item Repeat the second step until the number of gates is $d_{N,L} = \begin{pmatrix}	L \\ N \end{pmatrix}$.
\item Finally, fix two of the parameters to bring the count down to $2d_{N,L} - 2$ real parameters.
\end{itemize}

The algorithm generates a brick-wall variational quantum circuit that conserves the total particle number. While the circuit presented can be said to be optimal in the number $d_{N,L}$ of basic gates, we however discovered that it is possible to construct the desired variational circuit with $(d_{N,L}-1)$ number of basic parameterized gates, but with the aid of additional swap gates, and so the resulting circuit is sub-optimal. The basic underlying idea depends on the couplings of the qubits. We shall illustrate this fact below with the example of a lattice of $4$ sites with $2$ particles. This example will serve both to illustrate this fact and to provide a concrete example of the algorithm outlined above.

\subsection{Qubit coupling matters}
Here, we show that $(d_{N,L} - 1)$ number of the basic parameterized gates $A(\theta, \phi)$ or $B(\theta, \phi)$ can be used to build variational quantum circuits that can span the subspace $\mathcal{H}_{N,L}$, albeit at the expense of introducing addition swap operations. Therefore, it implies that a VQC with just the right number of parameters can be constructed directly, as opposed to the method used in Ref.~\cite{gard2020efficient} that we mentioned previously. 

\begin{figure}
\Qcircuit @C=0.25em @R=0.4em{
\ket{0} ~ &  & \qw & \multigate{1}{A(\theta_1, \phi_1)} & \qw & \multigate{1}{A(\theta_4, \phi_4)} & \qw & \qw \\
\ket{0} ~ &   & \gate{X} & \ghost{A(\theta_1, \phi_1)} & \multigate{1}{A(\theta_3, \phi_1)} & \ghost{A(\theta_1, \phi_1)} & \multigate{1}{A(\theta_6, \phi_6)} & \qw \\
\ket{0} ~ &   &  \gate{X} & \multigate{1}{A(\theta_2, \phi_1)} & \ghost{A(\theta_1, \phi_1)} & \multigate{1}{A(\theta_5, \phi_5)} & \ghost{A(\theta_1, \phi_1)} & \qw \\
\ket{0} ~ &   &  \qw & \ghost{A(\theta_1, \phi_1)} & \qw & \ghost{A(\theta_1, \phi_1)} & \qw & \qw
}
\caption{The symmetric variational quantum circuit for the example of a lattice of $2$ particles on $4$ sites, constructed according to the algorithm given in Sec.~\ref{Sec:U1 symmetric quantum circuit}. Two of the parameters have been fixed to reduce the parameter counts to 10.}
\label{Fig:QC2PtsOn4Sites_1}
\end{figure}
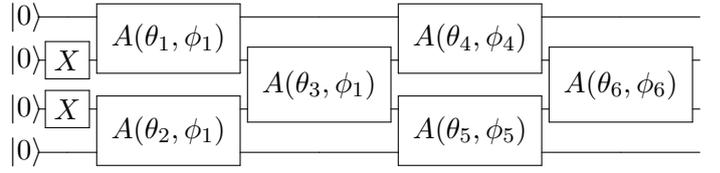

We use the example of 2 particles on 4 sites, with the subspace $\mathcal{H}_{2,4}$. The dimension of the subspace $\mathcal{H}_{2,4}$ is 6. A general state in this subspace can be parameterized with only 10 real numbers, if the state is normalized and an irrelevant global phase is discarded. Therefore, it should be possible to create a VQC for this example with only $5$ basic parameterized gates, e.g. A-gates. But according to the algorithm given in Sec.~\ref{Sec:U1 symmetric quantum circuit}, the ansatz circuit for this system is generated with $6$ basic gates as shown in Fig.~\ref{Fig:QC2PtsOn4Sites_1}. This circuit was confirmed numerically, using a fidelity test, to span the relevant subspace. To gain a better insight into this circuit, we examine the underlying couplings of the qubits, and we eventually discovered how to construct the VQC with only $5$ parameterized gates, but with additional swap operations.

\begin{figure}
    \centering
    \includegraphics{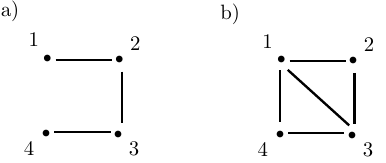}
    \caption{The underlying couplings of qubits initiated by the symmetric variational circuits above for the example of a lattice of $4$ sites with 2 particles. In subfigure a), there are only three couplings between the qubits, which are repeated twice, giving the $6$ basic gates in the corresponding circuit in Fig.~\ref{Fig:QC2PtsOn4Sites_1}. In subfigure b), more qubits are connected. There are five couplings and no bond is repeated.}
    \label{Fig:Qubit couplings}
\end{figure}

\begin{figure*}
\centering
\Qcircuit @C=1em @R=0.75em{
\ket{0} ~ &  & \qw  & \multigate{1}{A(\theta_1, \phi_1)} & \qw  & \qw & \qswap  & \qw & \qw & \qw & \qw & \qw  & \qw & \qw & \qswap & \qw & \qw \\
\ket{0} ~ &  & \gate{X} & \ghost{A(\theta_1, \phi_1)} & \multigate{1}{A(\theta_3, \phi_3)} & \qw & \qswap \qwx & \qw & \qswap & \qw & \multigate{1}{A(\theta_5, \phi_5)} & \qw  & \qswap & \qw & \qswap \qwx & \qw & \qw \\
\ket{0} ~ &  & \gate{X} & \multigate{1}{A(\theta_2, \phi_2)} & \ghost{A(\theta_1, \phi_3)} & \qw & \qw & \qw & \qswap \qwx & \multigate{1}{A(\theta_4, \phi_4)} & \ghost{A(\theta_5, \phi_5)} & \qw & \qswap \qwx & \qw & \qw & \qw & \qw \\
\ket{0} ~&  & \qw & \ghost{A(\theta_1, \phi_2)} & \qw & \qw & \qw & \qw & \qw & \ghost{A(\theta_4, \phi_4)} & \qw  & \qw & \qw & \qw & \qw & \qw 
}
\caption{Variational quantum circuit for a lattice with 2 particles on 4 sites. Note that this circuit is constructed with only 5 basic parameterized gates $A(\theta, \phi)$, though with additional swap operations. }
\label{Fig:QC2PtsOn4Sites_2}
\end{figure*}
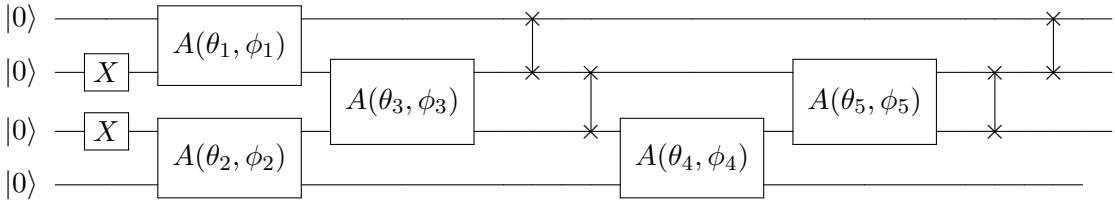

First, we number the qubits in the circuit (see Fig.~\ref{Fig:QC2PtsOn4Sites_1}) from top to bottom as $1,2,3,4$. We depict the couplings of the qubits as contained in the reference circuit as:
\begin{equation}
    (12)34, 12(34), 1(23)4, (12)34, 12(34), 1(23)4,
\end{equation}
where each grouping in brackets introduces a basic gate $A$ with two new parameters, $\theta_k, \phi_k$. The couplings can be summarized with Fig.~\ref{Fig:Qubit couplings}(a). It is seen that there is no direct connection between qubits $(1,3)$, $(2,4)$ and $(1,4)$. We imagined that this is the reason why a VQC constructed with only a $5$ parameterized gate according to the algorithm above would not work. We discovered that a different circuit where swaps are used to create more connectivity did work. The swap operations are applied to bring the qubits close, apply the parameterized gates on them, and then apply reverse swap operations to restore the qubits to their original positions. We discovered that the couplings below work:
\begin{equation}
    (12)34, 12(34), 1(23)4, \underline{(1)23(4)}, \underline{(1)2(3)}4,
\end{equation}
where in addition to the bracketed neighbouring qubits, $\underline{(1)23(4)}$ means qubits $1$ and $4$ are brought to couple and for $\underline{(1)2(3)}4$, qubits $1$ and $3$ are also brought to couple. 
The couplings are summarized with the diagram in Fig.~\ref{Fig:Qubit couplings}(b). Note that there are only 5 couplings in this case, and each introduces 2 parameters, therefore there are a total of 10 parameters. The circuit corresponding to this coupling is shown in Fig.~\ref{Fig:QC2PtsOn4Sites_2}. Note that this circuit uses only five parameterized gate $A(\theta_k, \phi_k)$ for $1 \leq k \leq 5$, in additional to some swap gates to bring qubits $(1,4)$ and $(1,3)$ to interact. Another coupling that was discovered to work similarly to Fig.~\ref{Fig:Qubit couplings} is to couple qubits 2 and 4 instead of qubits 1 and 3. These circuits were confirmed numerically to span the subspace $\mathcal{H}_{2,4}$. (For all the circuits above, the B gates can also be used in place of the A gates.)

The important note from this example is that to span a relevant subspace, the underlying couplings of the qubits matter. Although the circuit shown in Fig.~\ref{Fig:QC2PtsOn4Sites_2} uses a lesser number of parameterized gates than that in Fig.~\ref{Fig:QC2PtsOn4Sites_1}, it also uses more swap gates, and this may become prohibitive for longer lattices with more particles. {To make this more concrete, we can examine the ``complexity'' of the two circuits in terms of the number of CNOT gates required. We do this for circuits comprising just the $A(\theta, \phi)$ gate. (A similar conclusion is possible for the $B(\theta, \phi)$ gate.) The circuit in Fig.~\ref{Fig:QC2PtsOn4Sites_1} contains $18$ CNOT gates, where each A gate has $3$ CNOTs (see Eq.~\eqref{Eq:A gate}). While the circuit in Fig.~\ref{Fig:QC2PtsOn4Sites_2} contains $27$ CNOTs, where each swap requires $3$ CNOTs \cite{vatan2004optimal}. Therefore, since the circuit in Fig.~\ref{Fig:QC2PtsOn4Sites_1} is shorter in terms of the number of two-qubit gates, the error is smaller and hence should be more preferred. Moreover, as confirmed numerically through fidelity and energy minimization, the circuit in Fig.~\ref{Fig:QC2PtsOn4Sites_2} obtained with five basic gates and additional swap gates perform worse than the one with the six basic gates (without swaps). Therefore, the symmetric VQCs constructed using the algorithm outlined in Sec.~\ref{Sec:U1 symmetric quantum circuit} is preferred. But please note that this argument is not universal. It is made with respect to the qubit configuration shown in Fig.~\ref{Fig:Qubit couplings}. It is expected that the gate count would be different for a real device with connectivity different from the ones shown in Fig.~\ref{Fig:Qubit couplings}.}

\section{Test Results for Heisenberg Model}
\label{Sec:Test Model}
{As with proposing any new ansatz, we now test the effectiveness of our circuits, by using them to learn the ground state of the Heisenberg XXZ model. Firstly, we confirmed numerically that the circuits span the symmetric subspace through a fidelity test. Secondly, through energy minimization of the XXZ model, we compare the effectiveness of the two circuits, $\mathcal{C}_A$ and $\mathcal{C}_B$. Our numerical experiments were carried out on a quantum simulator implemented with Qiskit.\cite{Qiskit} }

\subsection{Fidelity of circuits}
We numerically confirmed that the aforementioned proposed variational circuits spanned the relevant subspace, by maximizing the fidelity of trial states $\ket{\Psi_i}$ against random states $\ket{\phi_i}$ chosen according to Haar measure within the subspace. For pure states, the fidelity can be written as
\begin{equation}
F = \frac{1}{N} \sum_{i=1}^N |\langle \phi_i \ket{\Psi_i}|^2, 
\end{equation}
where $\ket{\phi_i}$ are random states chosen within the relevant subspace. The circuits for the example of a lattice with 2 particles on 4 sites were confirmed to achieve the maximum fidelity of one, for both the circuits with 6 parameterized gates in Fig.~\ref{Fig:QC2PtsOn4Sites_1} (with either A- or B- gate) and also the circuit with 5 paramterized gates with additional swaps as in Fig.~\ref{Fig:QC2PtsOn4Sites_2}. For all examples, the variational circuits constructed according to the general algorithm given in Sec.~\ref{Sec:U1 symmetric quantum circuit} all achieved the maximum fidelity value.

\subsection{Energy minimization of XXZ Hamiltonian}
After the variational circuits were tested and confirmed to achieve maximum fidelity, we then used the circuits to obtain the ground state of the Heisenberg XXZ model using the variational quantum eigensolver (VQE) algorithm. 

The Hamiltonian for the XXZ model is
\begin{equation}
\label{Eq:XXZModel}
    H = \sum_{i=1}^L \left( X_i X_{i+1} + Y_i Y_{i+1} + \gamma Z_i Z_{i+1} \right),
\end{equation}
where $X,Y,Z$ are the usual Pauli matrices $\sigma^x, \sigma^y, \sigma^z$. This model conserves total magnetization $M = \sum Z_i$, which means that $[H,M]=0$. {There is a mapping between the XXZ model and the hardcore Bose-Hubbard model---a model with conserved total particle number. A variational quantum circuit for a tight-binding problem can therefore be tested using a quantum magnetism problem. This mapping is shown in Appendix \ref{App:XXZ2BHM}.}

\begin{figure}
    \centering
    \subfloat[$4$ sites]{\includegraphics[width=0.45\textwidth]{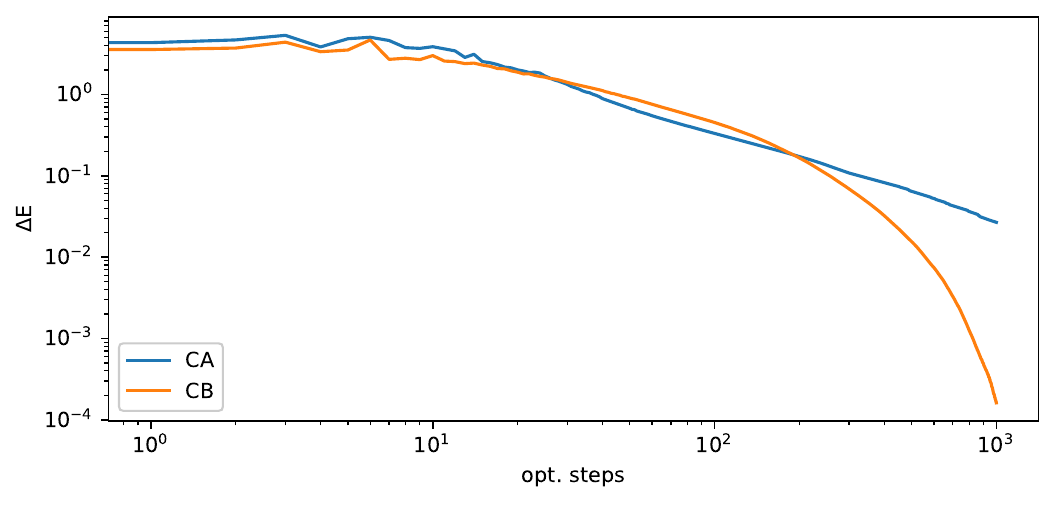}}
    
    \subfloat[$6$ sites]{\includegraphics[width=0.45\textwidth]{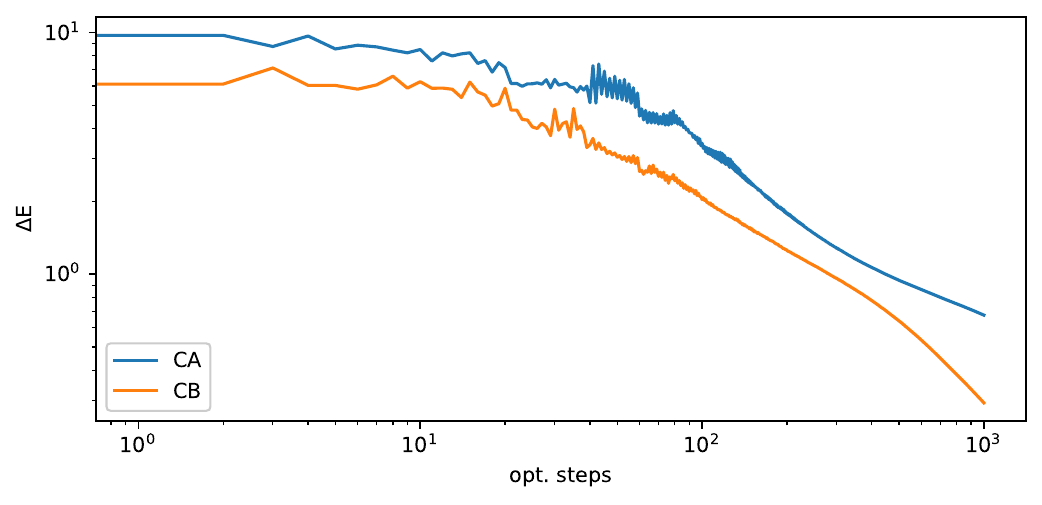}}
\caption{Results of the noiseless simulations of XXZ model for lattices with $4$ and $6$ sites. The plots show the energy difference $\Delta E$ against the number of optimization (opt.) steps. The energy difference $\Delta E$ is the difference between the numerical ground state energy and the exact ground state energy. The reference ground state energies are: $E_0 = -6.4641$ for $4$ sites, and $E_0 = -9.97431$  for $6$ sites.}  
\label{Fig:Noiseless simulation}
\end{figure}

We performed optimization of the XXZ Hamiltonian (with $\gamma = 1$) with the two variational circuits $C_A$ and $C_B$. {The ground state is obtained by minimizing the energy cost function
\begin{equation}
    E = \bra{\psi} H \ket{\psi},
\end{equation}
where the state $\ket{\psi}$ corresponds to its circuit ansatz. The ground state of the XXZ Hamiltonian $H$ has zero magnetization, $M=0$, which corresponds to the half-filling state of the Bose-Hubbard model. }

The optimization was implemented with IBM Qiskit, with both noisy and noiseless simulators. For the noisy simulation, we use the noise model that is saved in Qiskit Terra, which was extracted from a real IBM quantum device.

The numerical experiments were conducted for small lattices of $4$ and $6$ sites. The estimate of the energy expectation value was computed with 1024 shots. We used the COBYLA classical optimizer, with its default maximum number of iterations, 1000 (i.e. optimization steps). The optimization procedure generally proceeds by starting the ansatz circuit with some initial parameter values and then using the classical optimizer to update the values, which is then used for the next iteration. These steps are repeated until the simulation ends. 

For a fair comparison between the two circuits, we start both circuits with the same initial parameter values. But in order to eliminate any bias related to initialization, we ran the optimization over many trials and averaged the results. We performed $500$ trials.

\begin{figure}
    \centering
    \subfloat[$4$ sites]{\includegraphics[width=0.45\textwidth]{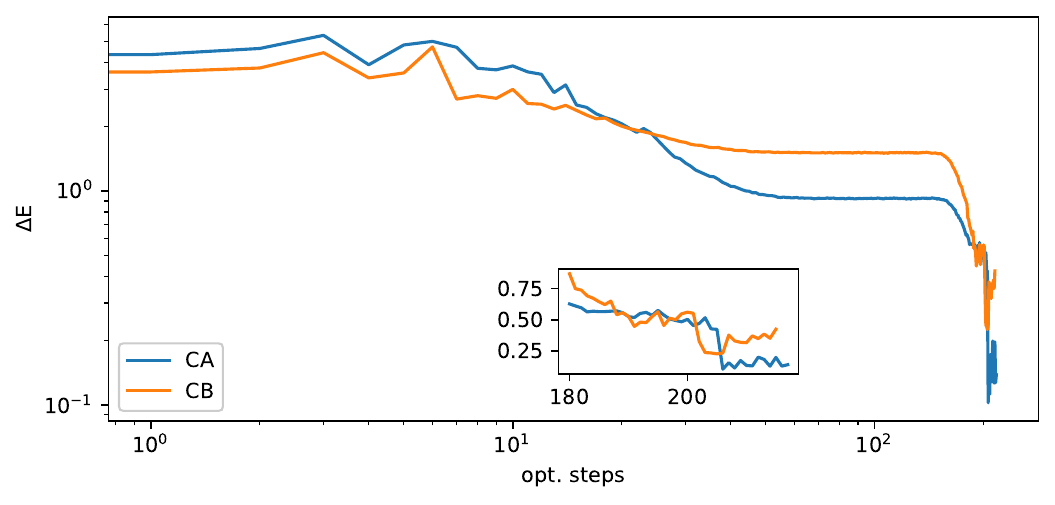}}
    
    \subfloat[$6$ sites]{\includegraphics[width=0.45\textwidth]{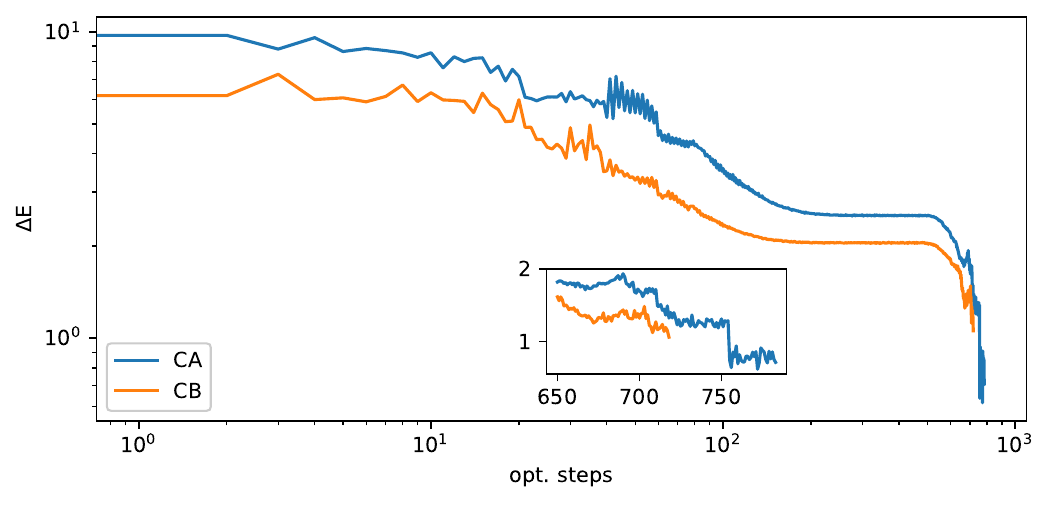}}
\caption{Results of the noisy simulations for lattices of $4$ and $6$ sites. The plots show the energy difference $\Delta E$ against the number of optimization (opt.) steps. The energy difference $\Delta E$ is the difference between the numerical lowest energy and the exact ground state energy.}  
\label{Fig:Noisy simulation}
\end{figure}

The results for the noiseless simulations are presented in Fig.~\ref{Fig:Noiseless simulation}, and the ones for the noisy simlations are presented in Fig.~\ref{Fig:Noisy simulation}. We explain these results below.

For the noiseless simulations (Fig.~\ref{Fig:Noiseless simulation}), on average, the circuit $C_B$ performed significantly better than circuit $C_A$ for the two lattice sizes considered. 

For the noisy case, in the limit of small optimization steps, say around two orders of magnitude, circuit $C_A$ appears to  perform better than $C_B$ for the $4$-site lattice, while the reverse is the case for the $6$-site lattice. I do not yet know the reason for this  behavior. However, in the long-time limit, simulations with circuit $C_ A$ converged better than circuit $C_B$, for both lattices as shown in the insets of Fig.~\ref{Fig:Noisy simulation}. 

We have not applied any error mitigation techniques in this work. The noisy simulations for both circuits performed worse than their noiseless counterparts, as expected. 

In summary, the variational circuits comprising of $B$-type gates may continue to perform better than $A$-type gates in the noiseless case even for larger lattices, while for noisy simulations, more and larger examples would have to be investigated before a solid, over-arching conclusion can be made.

\section{Conclusion}
\label{Sec:Conclusion}
In this paper, using ideas from representation theory we have shown how to build particle-conserving variational quantum gates using an approach that we referred to as ``bottom-to-top." In our approach, we derive individually all the elementary gates needed to construct particle-conserving gates. We then composed them together to build symmetric gates, which in turn we compose to build particle-conserving variational quantum circuits. Our approach is different from the ``top-to-bottom'' approach that has been considered previously in the literature, where a two-qubit general unitary matrix is constrained to be particle-conserving, and this is then decomposed into elementary quantum gates.

We derived explicitly two types of particle-conserving gates, which were correspondingly used to construct two different VQC ans\"{a}tze. We tested the symmetric VQCs with the Heisenberg XXZ model on a quantum simulator with and without noise. For the system sizes examined, in the noiseless cases, we obtained very good estimates of the ground state energy, while in the noisy cases and without error mitigation, the results were poorer, as should be expected. 

As a final remark, the methods exposed in this paper for constructing particle-conserving quantum gates should be applicable to other cases of quantum symmetries, including Abelian and non-Abelian ones. This will be investigated further in the future.

\section{Acknowledgement}
I thank Javier Osca Cotarelo for the preliminary code implementation in Qiskit at the initial stage of this project. {I also thank Prof. Jiri Vala for the suggestions that helped to improve the paper.} This work is supported with funding from Enterprise Ireland’s DTIF programme of the Department of Business, Enterprise, and Innovation, project QCoIr Quantum Computing in Ireland: A Software Platform for Multiple Qubit Technologies No. DT 2019 0090B.

\appendix

\section{Mapping XXZ Heisenberg Model to Bose-Hubbard Model}
\label{App:XXZ2BHM}
In this section, we show the mapping of the XXZ Heisenberg model to the hardcore Bose-Hubbard model. To this end, we use the following maps between Pauli matrices and (hardcore) bosonic second-quantized operators $a, a^{\dagger}$,
\begin{align}
\label{Eq:Pauli Second-quantized maps}
    X & = a + a^{\dagger} \nonumber \\
    Y & = -i(a - a^{\dagger})  \\
    Z & = 1 - 2n \nonumber,
\end{align}
where $a$ ($a^{\dagger}$) is a hardcore annihilation (creation) operator. Recall that the operators satisfy the following relations:
\begin{align}
    & \left[ a_i, a_j^{\dagger} \right] = \delta_{ij}, \\
    & a_i^2 = \left(a_i^{\dagger} \right)^2 = 0, \\
    & n_i = a_i^{\dagger} a_i,
\end{align}
where the second relation imposes the hardcore constraint. This implies that any site $i$ cannot have more than one particle. With a minor algebra, the XXZ Hamiltonian in Eq.~\eqref{Eq:XXZModel} can be written, up to some irrelevant terms, as 
\begin{equation}
\label{Eq: Bose-Hubbard model}
    H_{\mathrm{BH}} = \sum_i \left( a_i^{\dagger} a_{i+1} + a_{1+1}^{\dagger}a_i + \Delta n_i n_{1+1} \right),
\end{equation}
where $\Delta = 2\gamma$. 

This Hamiltonian $H_{\mathrm{BH}}$ can be checked to commute with the total particle number operator $N = \sum n_i$. Therefore, the total particle number is conserved. Furthermore, the particle number conservation of this Hamiltonian is related to the conservation of magnetization of the XXZ model. Indeed, using the third map in Eq.~\eqref{Eq:Pauli Second-quantized maps}, $\hat{M} = \sum_i Z_i = 1-2N$. So, in particular half filling, i.e. $N=L/2$ of the Bose-Hubbard model corresponds to the zero magnetization, $M=0$, of the XXZ model. Therefore, particle-conserving VQCs built for a tight-binding-like problem can also be used to simulate the Heisenberg XXZ magnetism model.

\bibliographystyle{plain}
\bibliography{References}

\end{document}